\documentclass[12pt]{elsart}
\usepackage{amsmath}
\usepackage{afterpage}
\usepackage{psfig}

\begin{document}  

\begin{frontmatter}
\title{Chance and Necessity in Evolution:\\Lessons from
RNA\thanksref{CNLS}}
\thanks[CNLS]{Submitted to {\it Physica D\/}, Special
Issue on ``Predictability: Quantifying Uncertainty in Models of
Complex Phenomena''}

\author{Peter Schuster} and
\author{Walter Fontana}

\address{Santa Fe Institute, 1399 Hyde Park Road,
Santa Fe, NM 87501 USA}
\address{Institut f\"ur Theoretische Chemie und
Strahlenchemie, Universit\"at Wien, W\"ahringerstra\ss e 17,
A-1090~Wien, Austria}

\begin{abstract}
The RNA folding map, understood as the relationship between sequences 
and secondary structures or {\it shapes}\/, exhibits robust statistical
properties summarized by three notions: (1) the notion of a
{\it typical shape}\/ (that among all sequences of fixed length 
certain shapes are realized much more frequently than others),
(2) the notion of {\it shape space covering}\/ (that 
all typical shapes are realized in a small neighborhood of any random 
sequence), and (3) the notion of a {\it neutral network}\/ (that 
sequences folding into the same typical shape form networks that 
percolate through sequence space).

The concept of a neutral network is particularly illuminating.
Neutral networks loosen the requirements on the mutation rate for
selection to remain effective. What needs to be preserved in a
population is not a particular sequence, but rather a shape. This
mandates a reformulation of the original (genotypic) error threshold
in terms of a phenotypic error threshold confirming the intuition
that more errors can be tolerated at higher
degrees of neutrality.

With regard to adaptation, neutrality has two seemingly contradictory
effects: It acts as a buffer against mutations ensuring that a
phenotype is preserved. Yet it is deeply enabling, because it permits
evolutionary change to occur by allowing the sequence context to vary
silently until a single point mutation can become phenotypically
consequential.  Neutrality also influences predictability of adaptive
trajectories in seemingly contradictory ways. On the one hand it
increases the uncertainty of their genotypic trace. At the same time
neutrality structures the access from one shape to another, thereby
inducing a topology among RNA shapes which permits a distinction
between continuous and discontinuous shape transformations. To the
extent that adaptive trajectories must undergo such transformations,
their phenotypic trace becomes more predictable.
\end{abstract}

\end{frontmatter}

\section{Introduction}

In his provoking classic, {\em Chance and Necessity\/}
\cite{monod:71}, Jacques Monod expressed the belief that a
``universal theory'', despite anticipating the appearance of certain
{\em classes\/} of objects (such as galaxies, planetary systems,
molecules, atoms, and the like), would not be able to account for the
biosphere. The biosphere, Monod says, does not contain a predictable
class of objects. Essential unpredictability from first principles
does not imply that the biosphere is not explicable through these
principles: biological objects have, in Monod's words, ``no obligation
to exist, but they have the right to'' (\cite{monod:71}, p. 44).  From
a more rigorous point of view, the problem of predicting the class of
objects that are outcomes of biological evolution, that is, species of
organisms, is ill-posed as long as we lack a formal specification of
this class \cite{fontana:94}. A judgement on Monod's position must,
therefore, remain open for the time being.

While the selection pressures of natural evolution arise endogenously,
artificial evolution allows to specify desired outcomes implicitly by
fixing those pressures exogenously. For example, when RNA molecules
are being intentionally evolved in the test tube to perform certain
functions or to bind certain targets \cite{ellington:94a}, an outcome
is being implicitly specified in advance. The evolutionary problem
then reduces to producing an actual RNA sequence folding into a shape
implementing some prespecified properties. This is an adaptative
process {\em within\/} a class of objects rather than the evolution
{\em of\/} a class of objects (the process which Monod believed to be
intrinsically unpredictable). Issues concerning the interplay of
chance and necessity become smaller in scope but also better defined
in the context of evolutionary adaptation. For example, is it
predictable whether a desired outcome can be attained?  And how
predictable are adaptive trajectories given some level of abstraction?

When considering the production of antibodies during an immune
response, Monod -- like many -- was puzzled by the effectiveness of
adaptation as a search engine driven by selection and mutation alone.
The puzzle about the effectiveness of adaptation is, as we shall
argue, only apparent.  It results from a misframing of the problem not
unlike in the so-called Levinthal paradox \cite{levinthal:69} of
protein folding, where the puzzle is (or rather, was) about how
proteins can fold into their native structure despite a
combinatorially large space of possible configurations.  The Levinthal
paradox has come to be recognized as resulting from a false dichotomy
\cite{dill:97}: either the protein has to make an extensive search
through its conformation space to find the lowest energy state (but
then it should not fold within observable times), or it folds ``down''
some energy path (but then it should get stuck in a suboptimal
trap). However, like protein folding, both the limitations and the
opportunties for evolutionary adaptation derive from specific features
of the ``landscape'' on which it occurs. The term landscape denotes
here a space of appropriately weighted pertinent configurations -
self-reproducing units and their fitness in the context of
evolutionary adaptation, or molecular conformations and their (free)
energy in the context of protein folding. Paths on such landscapes are
not equiprobable. The stochastic dynamics of both the adaptive process
(selection) and the folding process (least action) is guided by the
landscape structure and explores only a tiny fraction of the possible
configurations. This raises the issue about what exactly a ``folding
pathway'' or an ``evolutionary trajectory'' consist in, and how
well-defined they are. The ability to address the problem of
predictability in adaptation depends on such a characterization.

Exploring the structure of landscapes is a very active area of
empirical and theoretical research \cite{frauenfelder:97}. Despite the
similarity of certain questions, and despite the fact that both
processes crucially depend on special features of their landscapes,
the analogies between folding of an individual molecule and
evolutionary adaptation are limited. Landscapes underlying
evolutionary processes differ fundamentally from energy landscapes,
because biological entities are organized by a genotype-phenotype map.
Replication and mutation occur at the level of the genotype, but the
selective amplification of a genotype depends on the performance of
its phenotype. This dichotomy wouldn't be fundamental, if it were not
for the fact that the genotype-phenotype map is characteristically
many-to-one. The fact that many genotypes form the same phenotype
enables (as it turns out) evolutionary adaptation to be successful.
However, it also constitutes a major source of uncertainty in
evolutionary trajectories.

The split into genotype and phenotype implies two different notions of
``innovation'', and, hence, mandates care when speaking about
``trajectories''. Should an evolutionary trajectory refer to a
temporal succession of genotypes or of phenotypes?  And how are the
two related?  It is important to realize here that genotypic and
phenotypic innovation are not on an equal footing. A phenotype cannot
be modified directly, but only indirectly through variation of its
genotype.  This indirection means that phenotypic innovation is
mediated and, hence, biased by the genotype-phenotype map.  Genetic
mutations are random, but their consequences are far from random, as
they depend on the context in which they are expressed.  Assume, for
example, that all one-error mutants of a gene are equally
likely. Despite the absence of a bias at the genetic level, the
resultant protein shapes would, however, occur with biased
probabilities.  Innovation is locally isotropic for genotypes, but the
genotype-phenotype map channels phenotypic change along specific and
much fewer directions.  We should, therefore, expect regularities in
the genotype-phenotype map to reduce the ``phenotypic uncertainty'' of
evolutionary trajectories as compared to their ``genotypic
uncertainty''.

Imagine a game that is played on two boards, the g(enotype)-board and
the p(henotype)-board.  The game allows the player to make local moves
on the g-board only, while the actual pay-off is determined by moves
on the p-board. Some unknown machinery transduces the g-board-moves
into p-board-moves. Even if the g-board-moves were entirely random, an
observer of the p-board would pick up statistical regularities
reflecting the transducing mechanism.  Suppose further that an
``invisible hand'' holds the player's stone on the p-board fixed. The
transducing machine then acts back to confine the player to a subset
of moves on the g-board which are compatible with the stone's fixed
position on the p-board.  We call these moves neutral, as they don't
change pay-off.  The set of squares on the g-board that the player can
access by moving under this constraint we call a neutral
network. Imagine now the invisible hand softening its grip to allow
moves on the p-board that do not decrease pay-off (such as
selection). If none of the neighboring squares of the player's
position on the g-board yields a pay-off increase, the player won't be
stuck, as moves to neighboring neutral squares are still an
option. {\sl Their} neighbors might trigger pay-off increasing moves
on the p-board. As long as the stone on the p-board remains in a fixed
position, an observer of the p-board may conclude that nothing is
happening, and an observer of the g-board notes only ``more of the
same'' as the pay-off stays constant. The point, however, is that by
moving neutrally something {\em does\/} vary: the potential for
change.  Without neutrality the game would quickly become stuck in a
suboptimal trap, and the ``evolutionary Levinthal paradox'' would be a
real one.  However, with a sufficient degree of neutrality the very
notion of a ``trap'' looses its relevance, as ``lateral''
(fitness-neutral) moves change opportunities at no cost.

In this contribution we review work on the most realistic molecular
model for the genotype-phenotype dichotomy available to date.  RNA
unites both the genotypic and the phenotypic level in a single
molecular object.  The genotype is the sequence which can be
replicated in the test tube by suitable enzymes, and the phenotype is
the structure which can be subject to selection. We will summarize the
salient properties of the folding relation between sequences and
structures.  This sets the stage for discussing some central aspects
of evolutionary dynamics in models of evolving populations of RNA
molecules. We shall focus in particular on an emerging notion of
``evolutionary trajectory'' which is far from being fully understood
but which invites deeper investigation. This will prompt us to think
about what is and what is not predictable.

The main take-home should be an appreciation for the central and
apparently paradoxical role played by neutrality.  On the one hand,
neutrality (or redundancy, for that matter) is conservative, because
it buffers against inevitable mutations ensuring that ``nothing
happens'', that once attained success is not too easily lost. Yet at
the same time neutrality is deeply enabling, because it permits
evolutionary change to occur by allowing the sequence context to vary
until a mutation can become phenotypically consequential. In the
dynamics picture neutrality sets the stage for adaptation in jumps,
that is, intrinsic punctuation.  But neutrality does more: it endows
the set of phenotypes with a topological structure. A many-to-one map
induces equivalence classes of genotypes labelled by
phenotypes. Nearness (or accessibility) between phenotypes, then,
should be defined to reflect nearness between equivalence classes of
genotypes.  Once we possess a topology we can start thinking about
whether and when evolutionary trajectories are
``discontinuous''. While chance events will not allow to predict when
a discontinuity will happen, the topology enables the prediction of
what class of changes will be involved. We conclude by placing the
lessons learned from RNA into the larger perspective of evolution.

\section{RNA: an experimental and theoretical model system}
\label{RNAdef}

Understanding how notions like ``evolutionary path'' and
``continuity'' (or ``discontinuity'') are shaped by adaptive
landscapes, requires a model system that captures essential properties
of the relationship between genotype and phenotype. Since we don't
know these properties {\em a priori}, any theoretical model must have
a firm empirical grounding to begin with.  In addition to being
computationally tractable, the system should also be a laboratory
model of evolutionary adaptation. This excludes multicellular organisms, 
since their complexity does not permit at the present state of knowledge 
an empirical and theoretical tracking of how the genotype-phenotype
relation (i.e., development) plays out during an evolution experiment
in the laboratory.  Prokaryotic organisms don't undergo development
but the cellular metabolism is still too complex to be included
in a genotype-phenotype map. Settling for less than organismal complexity, 
we arrive at RNA as the simplest non-trivial molecular system fulfilling
``ideal model'' requirements.

What makes RNA unique is the simultaneous presence of both levels,
genotype and phenotype, in a single molecule. RNA molecules are
heteropolymers of (predominantly) four units called ribonucleotides.
Ribonucleotides have the ribose phosphate in common, but differ in
the base attached to the sugar. RNA molecules are represented as
sequences over a four letter alphabet, with each letter standing for a
particular base - {\bf A} for adenine, {\bf U} for uracil, {\bf C} for
cytosine, and {\bf G} for guanine.  Interactions mediated by hydrogen
bond patterns give rise to a stereoselective recognition between
specific pairs of bases - {\bf A}$\cdot${\bf U} and {\bf G}$\cdot${\bf
C}. This specific base pairing enables an RNA sequence to be copied 
{\it via\/} a complementary negative, and hence to function as a 
genotype. At the same time it enables segments of a sequence to pair 
with other segments within the same sequence, causing the sequence to 
fold back on itself into a structure. (In the formation of intramolecular
structure {\bf G}$\cdot${\bf U} pairs are possible as well.) This
structure mediates the chemical interactions of the sequence, and
hence constitutes its phenotype.

\begin{figure}[ht]
\centerline{\psfig{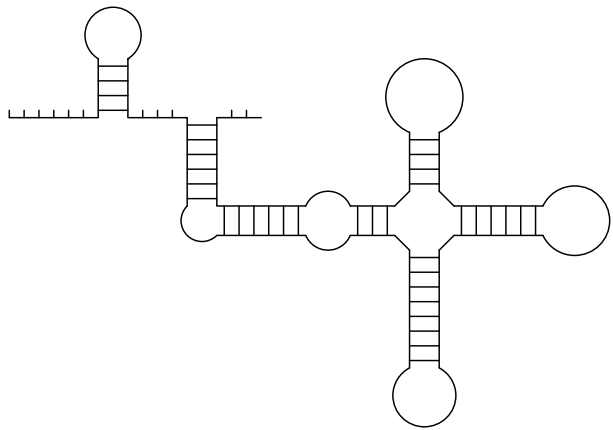}}
\caption[]{\label{secondar}{\small {\bf An RNA secondary structure graph.}
Unpaired positions not enclosed by base pairs, such as free ends or
links between independent structure modules, are called ``external''.
Here they are marked by ticks.}}
\end{figure}

``Structure'' can be seen at many levels of resolution, extending from
atomic coordinates to the mean radius of gyration.  This raises the
question of what constitutes a relevant level of structure resolution
for understanding adaptation. We take a pragmatic stance here, but we
shall return to this issue in conclusion. On the theoretical side we
require computability from the sequence, and on the empirical side we
require usefulness in interpreting molecular function and evolutionary
data. This leaves us with an empirically well established level of
resolution known as secondary structure.  The secondary structure of
an RNA molecule refers to a topology of binary contacts arising from
specific base pairing, rather than a geometry cast in terms of
coordinates and distances (Figure \ref{secondar}).  The driving force
behind secondary structure formation is the stacking between
contiguous base pairs.  The formation of an energetically favorable
paired (or double-stranded) region implies, however, the formation of
an energetically unfavorable loop. This ``frustrated'' energetics
leads to a vast combinatorics of stack and loop arrangements spanning
the structural repertoire of an individual RNA sequence.

A secondary structure can be conveniently discretized as a graph
representing the pattern of base pair contacts (Figure
\ref{secondar}). The nodes of the graph represent bases at positions
$i=1,\ldots,n$ along a sequence of length $n$. The set of edges
consists of two parts. One is common to all secondary structure
graphs, and represents the covalent backbone connecting node $i$ with
node $i+1$, $i=1,\ldots,n-1$. The other part is the secondary
structure proper, and consists of a set $P$ of edges $i\cdot j$,
$P=\{\;i\cdot j\;|\; i\ne j \text{ and } j\ne i+1\;\}$, representing
admissible hydrogen bonds between the bases at positions $i$ and
$j$. The set $P$ has to satisfy two conditions: (i) every edge in $P$
connects a node to at most one other node, and (ii) if both $i\cdot j$
and $k\cdot l$ are in $P$, then $i<k<j$ implies that $i<l<j$. Failure
to meet condition (ii) results in interactions (pseudoknots)
considered as belonging to the next - the tertiary - level of
structure. Both conditions distinguish RNA structure from protein
structure, in particular condition (i) which builds RNA secondary
structure exclusively from binary interactions.  We use a picture of
the graph as the visually most immediate representation of a secondary
structure. It proves convenient to also use a line oriented
representation, such as ``{\tt ((((.(((...))).(((...))).))))}'', where
a dot stands for an unpaired position, and a pair of matching
parentheses indicates positions that are paired with one another.

\begin{figure}
\centerline{\psfig{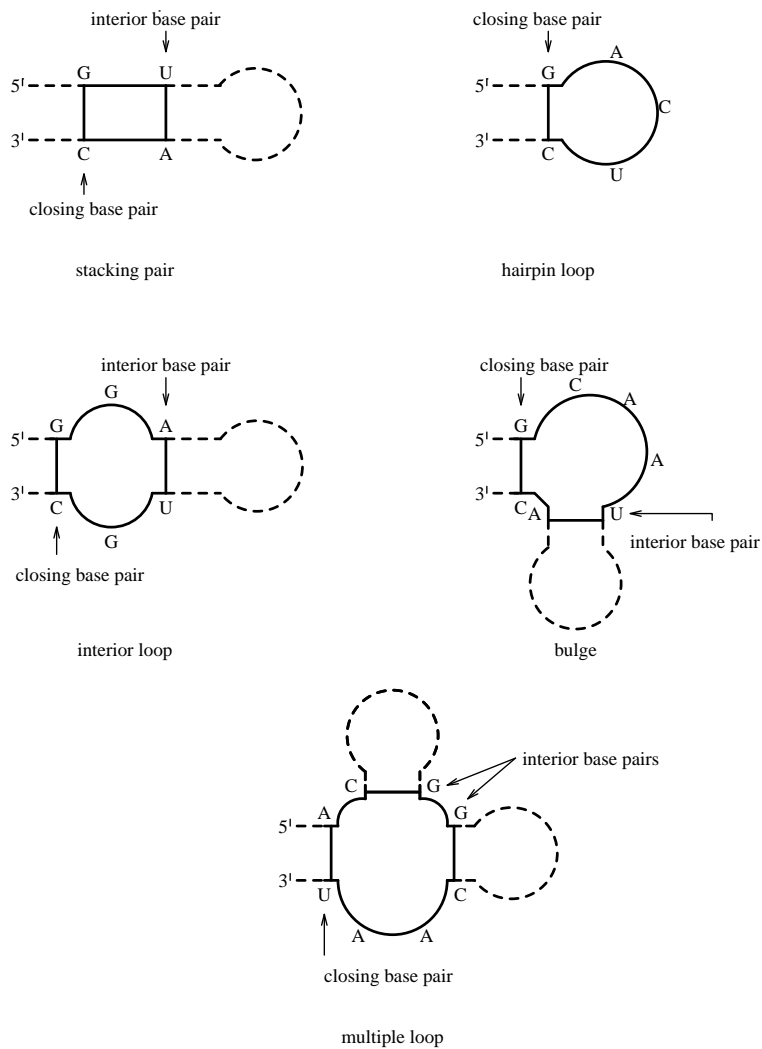}}
\caption[]{\label{elements} {\small {\bf Secondary structure elements.}}}
\end{figure}

Observe that the building blocks of a secondary structure are classes
of loops (Figure \ref{elements}): a hairpin loop is delimited by one
base pair which encloses a number of unpaired positions, a stack is
delimited by two base pairs and has no unpaired positions, while an
internal loop is delimited by two base pairs that enclose unpaired
positions. An internal loop is called a bulge, if either side has no
unpaired positions. Finally, the class of multiloops consists of loops
delimited by more than two base pairs. A position that does not belong
to any loop class is called external, such as free ends or joints
between loops (see Figure \ref{secondar}). The importance of these
loop classes derives from the reasonable assumption that the overall
energy of a secondary structure is the sum of its loop energies, and
from the fact that their free energies have been measured and
tabulated \cite{freier:86,turner:88,jaeger:89,he:91} as a function of
loop size and the nature of the delimiting base pairs.

Secondary structure graphs are formal combinatorial objects which can
be subject to mathematical treatment (they can be counted, for
instance).  Of particular interest are secondary structures possessing
some extremal property with respect to a given sequence, such as
minimizing the free energy. The theoretical importance of RNA as a
model system for sequence-structure relations in biopolymers lies in
the fact that structures of this kind can be computed by dynamic
programming
\cite{nussinov:78,waterman:78b,nussinov:80,zuker:81,zuker:84}. This
method produces a single structure that minimizes free energy.
Following an idea by Waterman \cite{waterman:85} we have recently
extended the standard RNA folding algorithm to compute all structures
within some energy range above the minimum free energy
\cite{wuchty:98}. However, for the present discussion the minimum free
energy structure will suffice, and we shall consider it to be the
phenotype of an RNA sequence.

The secondary structure is not just an utter abstraction, but it
provides both geometrically and thermodynamically a scaffold for the
tertiary structure. Its free energy accounts for a large share of the
overall free energy of the full structure. This linkage puts the
secondary structure in correspondence with functional properties of
the tertiary structure. Consequently, selection pressures (and hence
functional elements) become observable at the secondary structure
level in terms of conserved base pairs.

The extremely reduced amplification time and the minimal complexity of
the phenotype make RNA a tractable laboratory model.  RNA molecules
can be evolved in the test tube using a variety of techniques for
amplification, variation and selection. In fact, such experiments have
shown that evolutionary optimization of RNA properties in the test
tube occurs readily and effectively. Examples are the optimization of
replicative efficiency \cite{mills:67,spiegelman:71}, the production
of RNA molecules binding optimally to prespecified target molecules by
means of the SELEX technique \cite{ellington:90,tuerk:90},
evolutionary induced changes in the activity and specificity of
catalytic RNA molecules \cite{beaudry:92} (so-called ribozymes), and
the evolutionary design of ribozymes with new functions
\cite{bartel:93,ekland:95}.  Often the interpretation of such
experiments occurs by computing the secondary structure and placing it
in relation to molecular performance.

Until now the bulk of empirical interest has been directed at the
outcomes of evolutionary test tube experiments. However, we shall take
here outcomes for granted, and focus instead on dynamics and paths
toward outcomes. These aspects hold the key for understanding why (and
which) desired outcomes can be achieved at all.  An emerging
evolutionary technology will depend on a comprehensive dynamical
theory of molecular adaptation, much like chemical engineering depends
on chemical kinetics.

\section{Characteristic features of the RNA folding map}

The first computer experiments on the adaptive dynamics of replicating
and mutating RNA populations \cite{fontana:87} were very encouraging,
but they also made it clear that to understand adaptive dynamics, we
must first understand the features of the landscape induced by RNA
folding. The term ``landscape'' emphasizes the folding map as a
function from a {\it space\/} of sequences into a {\it space\/} of
structures (with possibly a further assignment of numerical values to
each structure).  The set of RNA sequences of length $n$ comprises
$4^n$ possible variations that are organized into a metric space by a
natural distance measure reflecting the allowed physical
interconversions of sequences. The Hamming distance is a natural
distance measure, if only point mutations are allowed. It counts the
number of positions in which two sequences differ, that is, the
minimum number of point mutations required to convert one sequence
into the other.  In the case of a two-letter alphabet this space is
the well-known $n$-dimensional hypercube.  The set of secondary
structures can also be made into a metric space.  Structure distance
functions are typically based on some notion of minimum edit cost for
transforming one structure into another, such as the Hamming distance
or the base pair distance \cite{hofacker:94} defined on the
dot/parentheses representation of secondary structures, or the
tree-edit-distance \cite{fontana:93b}.  If the structure space is
metric, it becomes possible to assess the ``ruggedness'' of the
landscape by autocorrelation functions \cite{fontana:93a}. However, it
is not always necessary or useful to think of the set of secondary
structures as a metric space. In section \ref{RNAtopo} we shall think
of it as a topological space.

Our interest is thus primarily aimed at extracting robust statistical
properties pertaining to the mapping as a whole. Although we use state
of the art algorithms \cite{zuker:81,hofacker:94,package} that are
routinely applied to predict the secondary structure of sequences
occurring naturally or having evolved in the laboratory, our main
focus is not the accurate prediction of a structure from a particular
sequence. (This would anyway require an integration with data from
phylogenetic comparison.)  We seek qualitative and generic features of
adaptive dynamics in RNA, and these should not depend on whether the
employed algorithms reproduce the fine details of the actual secondary
structure in a particular instance. Rather, we rely on the fact that
the employed algorithms are sufficiently mature that they correctly
capture the logic and basic energetics of constrained base-pair
optimization inherent in RNA folding. In fact, the generic features of
the folding map summarized below have been found to be insensitive to
the choice of structure formation criteria, such as minimizing free
energy, maximizing base pairing, or kinetic folding. They also are
numerically robust to variations in the set of empirical energy
parameters or the thermodynamic level of description (one minimum free
energy structure versus the Boltzmann ensemble for a given sequence)
\cite{tacker:96}. It is encouraging that similar properties have also
been recently discovered in lattice models of protein folding
\cite{goldstein:96,goldstein:97,li:96}. For the remainder of this
paper we shall - for the sake of brevity - refer to secondary
structures simply as ``shapes''.

A systematic study of the mapping from RNA sequences to shapes was
based on the statistics of appropriately chosen samples
\cite{fontana:93b,fontana:93a,schuster:94a} as well as on the
exhaustive folding of all sequences of a given length
\cite{gruener:96a,gruener:96b}. The regularities found depend on
two simple and fundamental facts. First, both the sequence and the
shape space are very high dimensional spaces (forget three-dimensional
caricatures), and, second, the sequence space is substantially larger
than the shape space. An upper bound along the lines of
\cite{stein:78} yields only $S_{n}=1.48 \times n^{-3/2}(1.85)^{n}$
shapes vis \`a vis $4^n$ sequences \cite{hofacker:99}. It is clear
that the mapping from sequences to shapes is significantly
many-to-one, even if the alphabet were binary. These, then, are the
major generic properties that were found:

Property 1 ({\it ``typical shapes''\/}) states that some shapes are
supported by significantly larger equivalence classes of sequences
(i.e., occur more frequently) than others. These relatively few
typical shapes are set apart from many rare shapes which can hardly
play a role in evolution.  The property of being ``typical'' is made
more precise by the observation that in the limit of long chains the
fraction of such shapes tends to zero (their number grows nevertheless
exponentially), while the fraction of sequences folding into them
tends to one.\footnote{This asymptotic condition for ``typical'' is
fulfilled by a whole class of definitions. A simple and
straightforward one is that of a so-called ``common shape'' which
refers to a shape formed by more sequences than the average given by
the number of sequences divided by the number of realized shapes
\cite{pks:234}.} A numerical example may help: The space of
{\bf GC}-only sequences of length $n=30$ contains $1.07\times10^{9}$
sequences folding into $218,820$ structures of which $22,718$
($=10.4\%$) classify as typical (in the sense of ``common'', see footnote
1). In this case $93.4\%$ of all sequences fold into these $10.4\%$
shapes.  Property 1 also implies that any statistical statement we
make about the folding map, and {\it a fortiori\/} about adaptive
dynamics, can only hold for the set of typical shapes.

\begin{figure}[ht]
\centerline{\psfig{figure=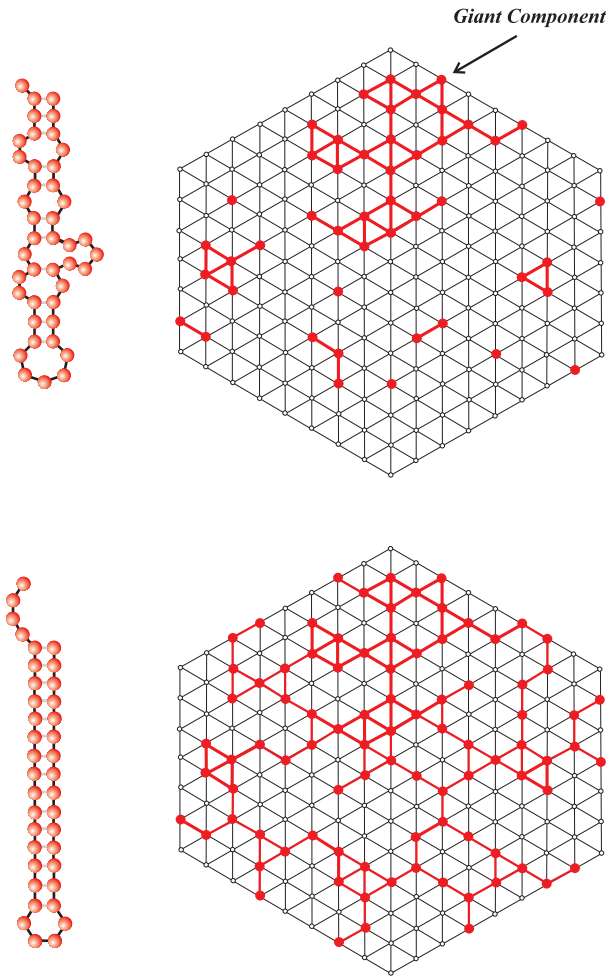,width=8cm}}
\caption[]{\label{neutnets}{\small {\bf Neutral networks in
sequence space.}  The lower half exhibits a typical structure with a
connected network of sequences folding into it. The network reaches
through sequence space. In contrast, the upper half shows an
``untypical'' or rare structure whose neutral sequences are far less 
in number. They are organized into a so-called giant component and many
small pockets. Connectivity of neutral networks depends on the mean
fraction of neutral neighbors ($\lambda$).}}
\vspace*{0.3cm}
\end{figure}

Property 2 ({\it ``neutral networks''\/}) is a statement about the
connectedness in sequence space of sequences folding into the same
shape.  Typical shapes are characterized by a high degree of
neutrality expressed as the average fraction of nearest neighbors of a
sequence possessing a typical shape that retain that shape. A large
enough degree of neutrality, expressed as the mean fraction $\lambda$
of neutral neighbors with Hamming distance one, leads to percolation
in sequence space, that is, to the existence of extended neutral
networks connecting sequences with the same shape by one or at most
two point mutations (Figure \ref{neutnets}).

\begin{figure}[ht]
\centerline{\psfig{figure=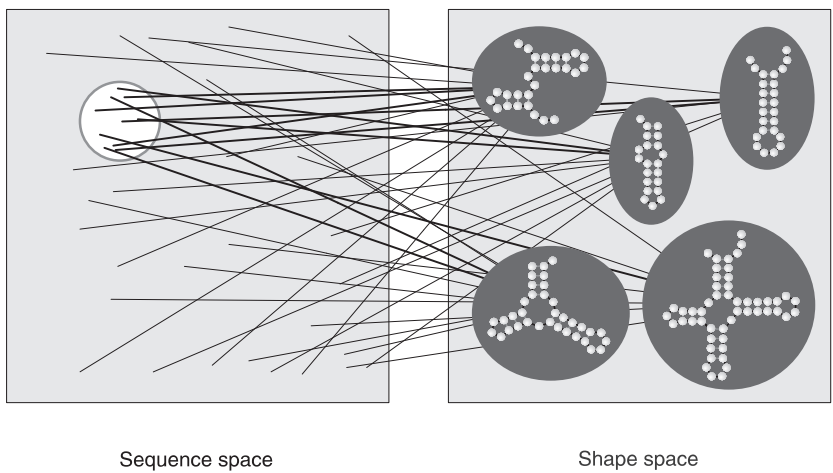,width=11cm}}
\caption[]{\label{covering}{\small {\bf RNA shape space covering.}
To find a sequence folding into a typical structure, only a relatively
small hyperspherical region around any random sequence needs to be
searched. The covering radius can be computed from properly chosen
samples of structures.}}
\vspace*{0.3cm}
\end{figure}

Property 3 ({\it ``shape space covering''\/}) is a statement about the
mutual entanglement of neutral networks belonging to different
shapes. All typical shapes are realized within a small neighborhood
(compared to sequence length) of any arbitrarily chosen sequence
(Figure \ref{covering}).  For example, 15 mutations are sufficient on
average for a chain of length $n=100$ to find at least one instance of
every typical shape.

These statistical properties, in particular neutral networks, have led
to a mathematical model based on percolation in random graphs
\cite{reidys:97}. There is no doubt that neutrality is an essential
feature of adaptive topographies. It is, therefore, crucial that
models of adaptive landscapes take neutrality into account, see for
example
\cite{derrida:91,gavrilets:97,newman:98,nimw98a,nimw98b,pks:248}.  We
shall explore some of its effects in the following two sections.

\section{Error propagation in genotype and phenotype populations}
\label{RNAdyn}

\begin{figure}[ht]
\vspace*{0.3cm}
\centerline{\psfig{figure=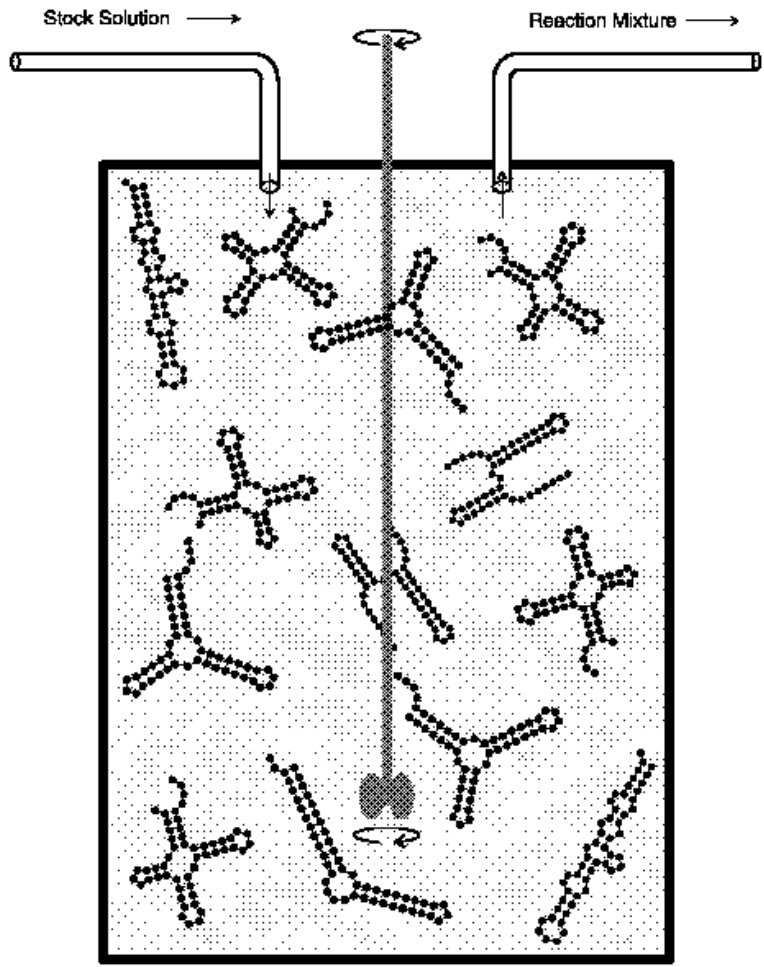,width=10cm}}
\caption[]{\label{chemostat}{\small {\bf Flow reactor.} The flow
reactor (continuously stirred tank reactor, CSTR) is a device for
recording chemical reactions with instantaneous replenishment of
consumed materials in continuous time. Like a chemostat for bacterial
cultures, it is used here to study the evolution of populations under
replication and mutation. The influx of stock solution, containing the
materials required for replication, is compensated by an unspecific
outflux of the reaction mixture which is kept homogeneous by
mechanical stirring. The flow is adjusted to yield an average of
$N\pm\sqrt N$ RNA molecules in the tank. In the computer experiments
described here the chain length $n$ of the molecules was kept constant
by restricting variation to point mutations. Parameters are the
population size $N$, the chain length $n$, and the mutation rate per
nucleotide and replication, $p$.}}
\vspace*{0.3cm}
\end{figure}

The evolutionary dynamics of molecules based on replication, mutation
and selection induced by a constant population size in a flow reactor
has been analyzed in terms of chemical reaction kinetics by Manfred
Eigen \cite{eigen:71}, and was further developed in subsequent studies
\cite{pks:78,eigen:89}.  Faithful copying and mutation
are treated on an equal footing by viewing the replication of an RNA
sequence as a branching reaction with many channels. In principle
every sequence can be obtained as a mutant from every sequence
(although the probabilities vary dramatically). The materials consumed
by RNA synthesis are replenished by a continuous flow in a reactor
resembling a chemostat for bacterial cultures (Figure
\ref{chemostat}).  The character of the dynamics of a sequence
population depends critically on two factors, the accuracy of
replication which governs the ``width'' of individual reaction
channels and the degree of neutrality - a property of the
genotype-phenotype map - which governs the decoupling of the
propagation of a given phenotype from the propagation of its
underlying genotypes. The presence of neutrality mandates a
distinction between a genotypic and a phenotypic error threshold
\cite{huynen:96a}.  The point of this section is to caricature the
essence of this distinction in a very simple way.  This will set the
stage for discussing trajectories as influenced by the RNA
genotype-phenotype map in section \ref{RNAtopo}.

Selection dynamics in sequence populations can be described by
ordinary differential equations. The variable $ x_i, 0\le x_i\le1$,
denotes the relative frequency of a sequence type (i.e., genotype)
I$_i$ of length $n$.  For the sake of brevity we shall overload the
word ``sequence'' to mean an individual instance or a type, depending
on context.  Since frequencies are normalized, we have $\sum  x_i=1$.
We further denote the population size with $N$, and the number of
possible sequences with $M=4^n$. Let the rate constants for replication
and degradation of sequence I$_i$ be $a_i$ and $d_i$,
respectively. The time dependence of the sequence distribution is then
described by the kinetic equations
\begin{equation}
\dot x_i\ =\ \Bigl(a_i Q_{ii}-d_i-\bar E(t)\Bigr) x_i\,+\,\sum_{j\neq
i} a_j Q_{ji}  x_j\ ,\quad i,j\,=\,1,\hdots,M\ .
\end{equation}
It has been assumed for simplicity that replication is direct, rather
than proceeding through a complementary negative, as the base pairing
rules would require (see \cite{eigen:71,eigen:89} for details).  The
width of the reaction channel from sequence I$_i$ to sequence I$_j$ is
given by the mutation matrix $Q\doteq\{Q_{ij};\,i,j=1,\ldots
M\}$. $Q_{ij}$ denotes the likelihood that a replication of sequence
I$_i$ yields sequence I$_j$, and the diagonal element $Q_{ii}$ is the
fraction of correct replicas synthesized on template I$_i$. To fulfill
$\sum \dot x_i = 0$, the reactor outflow $\bar E(t)$ matches exactly
the average productivity, $\bar E(t)=\sum_{i=1}
(a_i-d_i)\, x_i(t)$. When degradation is negligible, as
in the test tube evolution of RNA molecules or when all degradation
rates are essentially the same, $d_1\approx d_2\approx
\ldots\approx d_m\approx d$, degradation has no influence on
the selection dynamics and can be neglected.  In this case the
quantities which determine the selection dynamics are given by the
so-called value matrix: $W\doteq\{w_{ij}=a_i Q_{ij}\,;\,i,j=1,\ldots
M\}$, whose diagonal elements $w_{ii}$ were called selective values.

The selective value of sequence I$_i$ amounts to its fitness in the
case of vanishing mutational backflow, $\sum_{j\neq i} a_j Q_{ji}
 x_j\,<<\,a_i Q_{ii}\, x_i\,=\,w_{ii}\, x_i\,\, , \forall\,
i=1,\ldots,M$. Under these conditions the sequence I$_m$ with the
maximal selective value
\begin{equation}
w_{mm}\ =\ \max\,\{w_{ii}\,|\,i=1,\ldots,M\}\ , 
\end{equation}
dominates a population in selection equilibrium, and is called the
master sequence.

The term quasispecies was introduced for the stationary sequence
distribution, whose values $\bar x_i$ are computed as the solutions
of $\dot x_i=0;\,i=1,\ldots,M$, from the eigenvalue problem
\cite{thompson:74,jones:75}
\begin{equation}
\left(W\,-\,\omega\,I\right)\,\bar x\ =\ 0\ , \label{eigen}
\end{equation}
where $I$ denotes the unit matrix, $\omega$ an eigenvalue and
$\bar x=(\bar x_1,\bar x_2,\ldots,\bar x_M)$ the corresponding
eigenvector.

\subsection{Genotypic error threshold}

To understand the difference between genotypic and phenotypic error
thresholds, we consider the so-called single peak landscape which
assigns a higher replication rate to the master and identical values
to all others, $a_m=\sigma_m\cdot a$ and $a_i=a\, ,\forall\,i\neq m$,
where $\sigma_m$ is the superiority of the master sequence.  The
assumption behind the single peak landscape is to lump together all
mutants into a cloud with average fitness. The relative population
frequency of the cloud is simply $ x_c\,=\,\sum_{j\neq i}
 x_j\,=\, 1- x_m$, and the replication-mutation problem boils down to
an exercise in one variable, $ x_m$, the frequency of the master. The
approach has something in common with the mean field approximation
often used in physics, since the mutant cloud can be characterized by
``mean except the master'' properties. For example, the ``mean except
the master'' replication rate constant $\bar a\,=\,\sum_{j\neq m}
a_j x_j \big/ (1- x_m)$. The superiority then reads: $\sigma_m = a_m
\big/\bar a$.

Neglecting muational backflow we can readily compute the stationary
frequency of the master sequence in a ``zeroth-order'' approximation:
\begin{equation}
\bar x_m\ =\ \frac{a_m Q_{mm}\,-\bar a}{a_m\,-\,\bar a}\ =\ 
   \frac{\sigma_m Q_{mm}\,-\,1}{\sigma_m\,-\,1}\ .
\end{equation}
In this expression the master sequence vanishes at some finite
replication accuracy,
$Q_{mm}\bigm|_{\bar x_m=0}\,=\,Q_{min}\,=\,\sigma_m^{\ -1}$. A
non-zero stationary frequency of the master, that is, its indefinite
propagation, thus requires $Q_{mm}>Q_{min}$. This is the so-called
error threshold condition. It is illuminating to introduce a simple
model for the elements of the mutation matrix, called uniform error
rate approximation \cite{pks:78}. It assumes the existence of a per
nucleotide mutation probability $p$ that is independent of the nature
of a nucleotide and its position in the sequence. In terms of the
single digit accuracy $q=1-p$ (the mean fraction of correctly
incorporated nucleotides) the elements of the mutation matrix for a
sequence of length $n$ take the form:
\begin{equation}
Q_{ij}\ =\ q^n\,\left(\frac{1-q}{q}\right)^{d_{ij}}\ ,
\end{equation}
with $d_{ij}$ being the Hamming distance (i.e., the number of
mismatches) between two sequences I$_i$ and I$_j$. The critical
condition occurs at $q_{min}=1-p_{max}=\root n\of{\mathstrut
Q_{min}}=\sigma_m^{\ -1/n}$.  The replication accuracy of RNA viruses
was indeed found to occur near this critical value
\cite{domingo:96b,domingo:97a}.

To study stochastic features of the population dynamics around the
critical replication accuracy, the replication-mutation system has
been modeled by a multitype branching process \cite{pks:152}. The main
result of this study was the derivation of an expression for the
probability of survival to infinite time for the master sequence and
its mutants. In the regime of sufficiently accurate replication the
survival probability is non-zero and decreases with increasing error
rate. At the critical accuracy $q_{min}$ this probability becomes
zero, which implies that all molecular species currently in the
population will die out in finite time and new variants will replace
them.  This scenario is tantamount to migration of the population
through sequence space.

Further details of the population structure require consideration of
the mutational backflow (and thus a solution to the eigenvalue
problem, equation \ref{eigen}). The two eigenvectors associated with
the two largest eigenvalues at $q>q_{cr}$ represent the quasispecies
and the uniform distribution,
$\bar x_1=\bar x_2=\ldots=\bar x_M=1/M$, respectively.  These
eigenvectors show avoided crossing at the critical accuracy $q=q_{cr}$
\cite{pks:178}. Since the off-diagonal elements are small and decrease
further with increasing chain length $n$, the zeroth order
approximation to the frequency of the master is fairly good
($q_{min}\approx q_{cr}$), and the transition from the quasispecies to
the uniform distribution is sharp
\cite{eigen:89,pks:126}. Obviously, the deterministic approach
becomes meaningless at accuracies below threshold, when the uniform
distribution, $1/M=1/{4^n}$, implies less than a single molecule for
each variant. This is already the case for fairly small chain lengths
of $n>24$, when the population size required for the deterministic
approach ($N=4^n$) exceeds the feasible sizes realizable in test tube
evolution experiments ($N\approx10^{15}$).

\subsection{Phenotypic error threshold}

In the case of neutrality the superiority of the master becomes
$\sigma_m=1$, which pushes the error threshold towards absolute
replication accuracy, $q_{min}=1$, and the deterministic model must
fail. Any sequence on a neutral network is inevitably lost at non-zero
mutation rates, but the phenotype associated with that network may
nevertheless persist.  Neutrality, thus, shifts the focus from
genotypes to phenotypes, suggesting a treatment where phenotypes
become the relevant units by lumping together sequences with equal
phenotype.  Given $L$ possible phenotypes, we define new aggregate
variables, $\eta_{\alpha}$ ($\alpha=1,\ldots,L$), by collecting the
set of sequences with given phenotype $\alpha$ and replication rate
$a_{\alpha}$ (latin letters refer to sequences and greek ones to
phenotypes):
\begin{equation}
\eta_{\alpha}\ =\negthickspace\negthickspace
\sum_{\substack{\text{I$_i$ with}\\
\text{phenotype $\alpha$}}}
\negthickspace\negthickspace x_i\ ,
\end{equation}
with $\sum_{\alpha=1}^L\eta_{\alpha}=\sum_{i=1}^M  x_i=1$.  In
analogy to the previous treatment we call the phenotype $\mu$ with
maximal fitness $a_{\mu}$ the master phenoptype. Since we are heading
again for a zeroth-order approximation, we only consider the master
phenotype. Without loss of generality we index the sequence types
possessing the master phenotype from $1$ to $k$, so that
$\eta_{\mu}=\sum_{i=1}^k x_i$. This yields the kinetic equations for
the {\it set\/} of sequences with master phenotype:
\begin{equation}
\dot \eta_{\mu}\,=\,\sum_{i=1}^k \dot  x_i\,=\,\eta_{\mu}\,\bigl(
a_{\mu} Q_{kk} - \bar E\bigr)\,+\,
\sum_{i=1}^k \sum_{j\neq i} a_j Q_{ji}  x_j\ . \label{pheno-kin}
\end{equation}
The mean excess productivity of the population is, of course, independent 
of the choice of variables:
\begin{equation}
\bar E\,=\,\sum_{\alpha=1}^L a_{\alpha} \eta_{\alpha}\,=\,\sum_{i=1}^M
a_i  x_i\ .
\end{equation}
To derive a suitable expression for the phenotypic error threshold, we
split the mutational backflow into two contributions, (i) a mutational
backflow within the neutral network of the master phenotype and (ii) a
mutational backflow on to the master network from sequences not
belonging to it:
\begin{equation}
\sum_{i=1}^k \sum_{j\neq i} a_j Q_{ji}  x_j\,=\,\left\{
a_{\mu} \sum_{i=1}^k \sum_{j=1,j\neq i}^k Q_{ji}  x_j\right\} +
\left\{\sum_{i=1}^k \sum_{j=k+1}^M a_j Q_{ji}  x_j\right\}\ .
\end{equation}
We approximate the within-network backflow by assuming that a sequence
on the network has a constant fraction $\lambda_{\mu}$ of neutral
neighbors. We further assume equal mutation rates
($Q_{ji}=\bar{Q}_j;i,j=1,\ldots,k;i\neq j$) on the master network and
find:
\begin{eqnarray}
\sum_{i=1}^k \sum_{j=1,j\neq i}^k Q_{ji}  x_j\,&\approx&\,
\frac{\lambda_{\mu}(1-Q_{kk})}{k-1} \sum_{i=1}^k
\sum_{j=1,j\neq i}^k  x_j\,=\\
&=&\,\frac{\lambda_{\mu}(1-Q_{kk})}{k-1} 
\sum_{j=1,j\neq i}^k \sum_{i=1}^k  x_j\,=\,\lambda_{\mu}
(1-Q_{kk})\,\eta_{\mu}\ .
\end{eqnarray}
To keep things comparable, we make the same approximation as in the
genotypic error threshold and neglect mutational backflow from other
networks ($\eta_{\alpha},\alpha\neq\mu$) on to the master network. The
kinetic equation (\ref{pheno-kin}) for the master phenotype can now be
written as:
\begin{equation}
\dot \eta_{\mu}\,=\,\Bigl(a_{\mu}\tilde Q_{\mu\mu} - \bar E\Bigr)\eta_{\mu},
\end{equation}
where $\tilde Q_{\mu\mu}$ expresses an effective replication accuracy of the
master network as such:
\begin{equation}
\tilde Q_{\mu\mu}\,=\,Q_{kk}+\lambda_{\mu}(1-Q_{kk})\ .
\end{equation}

\begin{figure}[ht]
\centerline{\psfig{figure=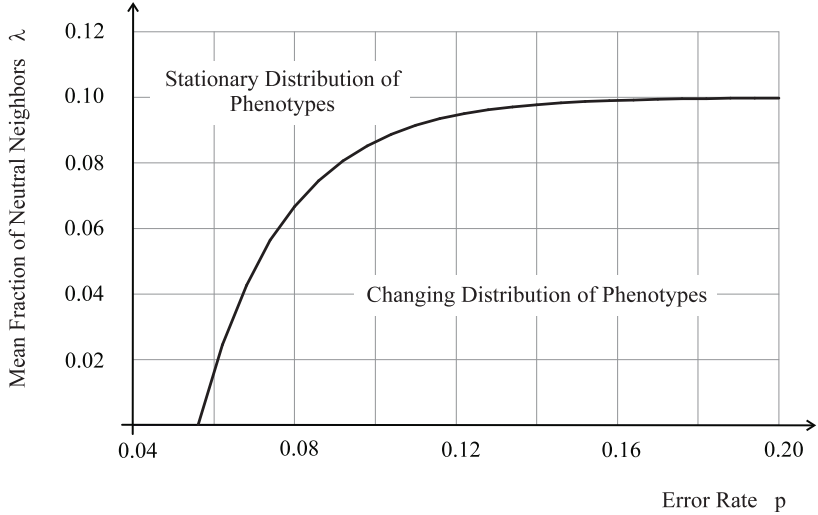,width=11cm}}
\caption[]{\label{phenothresh}{\small {\bf Phenotypic error threshold.}
The error threshold, $\tilde p_{\max}$, is shown as a function of the
error rate $p$ and the mean degree of neutrality, $\lambda$. The line
separates the domains of a stationary distriution of phenotypes and 
migrating populations. More erros can be tolerated at higher degrees
of neutrality.
}}
\vspace*{0.3cm}
\end{figure}

Proceeding in complete analogy with the derivation of the genotypic
error threshold, we find
\begin{equation}
\tilde Q_{min}\,=\,Q_{kk}+\lambda_{\mu}(1-Q_{kk})\,=\,\sigma_{\mu}^{-1}
\end{equation}
where $\sigma_{\mu}$ is the superiority of the master phenotype. The
uniform error rate model yields for the stationary frequency of the
master phenotype:
\begin{equation}
\bar\eta_{\mu}(p)\ =\ \frac{\sigma_{\mu}\,Q_{\mu\mu}(p)-1}
{\sigma_{\mu}-1}\ =\ \frac{(1-p)^n\,\sigma_{\mu}\,(1-\lambda_{\mu})\,+\,
\sigma_{\mu}\lambda_{\mu}\,-\,1}{\sigma_{\mu}-1}\ .
\end{equation}
The ``zeroth-order'' approximation for the phenotypic error threshold
($\bar\eta_{\mu}=0$) now becomes:
\begin{equation}
\tilde q_{min}\,=\,(1-\tilde p_{max})\,=\,
\left(\frac{1-\lambda_{\mu}\sigma_{\mu}}
{(1-\lambda_{\mu})\,\sigma_{\mu}}\right)^{1/n}\ .
\end{equation}
The function $q=\tilde q_{min}(n,\lambda_{\mu},\sigma_{\mu})$ is 
illustrated in Figure \ref{phenothresh}. In the limit of no neutrality,
$\lambda_{\mu}\to 0$, both phenotypic and genotypic error threshold
are the same, $\tilde q_{min}=\sigma_{\mu}^{-1/n}=\sigma_m^{-1/n}$. In the
limit of ``extensive'' neutrality,
$\lambda_{\mu}\to\sigma_{\mu}^{-1}$, the minimal replication accuracy
$\tilde q_{\min}$ approaches zero.  This means that when the degree of
neutrality exceeds the reciprocal superiority, the master phenotype is
never lost from the population, no matter what the mutation rate is.

This section's goal was to show that neutrality leads to populations
whose sequences drift randomly on the neutral network of the master
phenotype, or, for that matter, any temporary most fit phenotype
\cite{huynen:96a}.  Even if the sequence distribution drifts, the
phenotype is maintained. However, if the replication accuracy falls
below a critical minimum value - the phenotypic error threshold - the
population drifts through both sequence and shape space.
Interestingly, if the neutrality associated with a phenotype exceeds a
certain level, the phenotypic error threshold disappears, and that
phenotype cannot be lost from the population at any mutation rate.

Neutrality suggests a reformulation of Eigen's original kinetic
description and its stochastic versions in terms of neutral networks
as the units of aggregation. The switch in variables is complicated by
the fact that the sequence mutation matrix $Q_{ji}$ must be translated
into a ``mutation'' matrix $\tilde Q_{\alpha\beta}$ which refers to
transitions between neutral networks, i.e. phenotypes. This is where
we heavily glossed over in this section, since it requires an
understanding of the sense in which networks are ``adjacent'' to one
another.  One quite fruitful approach is to define prototype landscapes
\cite{reidys:97,nimw98a,nimw98b,nimw97a}, another is to analyze this
adjacency relation in the concrete case of RNA
\cite{fontana:98b}, and to extract regularities which may guide the
design of model landscapes.  We shall turn to this in the next
section.

\section{Adaptive paths and intrinsic punctuation}
\label{RNAtopo}

In this section we lay some groundwork for framing adaptive
trajectories in evolving RNA populations. To this end we must first
understand the organization of RNA shape space.

\subsection{The topology of RNA shape space}

We have seen in section \ref{RNAdyn} that neutrality loosens the
requirements on the replication accuracy for selection to remain
effective. This increases the uncertainity of adaptive trajectories
through sequence space, as populations with tiny sizes compared to the
number of all possible sequences undergo neutral diffusion.  At the
same time, however, adaptive trajectories are subject to quite
specific constraints. To see this informally, consider that for
phenotypes to adapt, transitions between neutral genotype networks
(that is, phenotypes) must occur. Some transitions are easy, others
not. If the replication accuracy is sufficiently high, so that most
mutants are one-error mutants, the degree to which a transition from
one network to another is easy depends on how much of their boundary
both have in common. This suggests defining a nearness relation
between phenotypes (a topology) based on the extent to which their
corresponding neutral networks are adjacent in genotype space
\cite{fontana:98b,fontana:98a}. Notice that a so defined nearness
relation is independent of a notion of similarity between phenotypes.
Transitions between neutral networks sharing a small fraction of their
boundary will, then, act as bottlenecks, and the uncertainty of
adaptive trajectories will be reduced to the extent that they must
pass such bottlenecks.  If we can characterize the set of possible
bottlenecks, we should be able to predict a {\it class\/} of phenotype
transformations that any adaptive trajectory must go through, although
we may not be able to predict the exact phenotypes involved or their
temporal succession. We shall call this class a class of ``way
points''.

In RNA, as in many biological situations, there are two mappings
involved which need to be kept distinct: the mapping from genotype to
phenotype and the mapping from phenotype to fitness. Both are
typically many-to-one and induce neutrality. However, the map from
phenotypes to fitness depends on the (exogenous or endogenous)
selection criteria, while at least in the case of RNA folding the
genotype to phenotype map does not. Because the class of phenotype
transformations that we are seeking is an intrinsic property of a
given genotype-phenotype relation, it constrains adaptive trajectories
in a fashion that holds under any (non-trivial) fitness assignment.

We shall now make these intuitions precise, and characterize the class
of way points for RNA secondary structures based on a statistical
analysis of the folding map.  We then juxtapose these results with
adaptive trajectories obtained from simulating a population of RNA
sequences that replicate and mutate in a flow reactor under selection.

A particular adaptive path consists of a temporal succession of
sequences and their associated shapes. We refer to the temporal series
of shapes as the phenotypic trace.  Whether a shape $\beta$ succeeds a
shape $\alpha$ will be strongly influenced by fitness. Yet for this to
be an issue at all $\beta$ must first occur somehow, that is, $\beta$
must be {\it accessible\/} from $\alpha$ by a mutation of $\alpha$'s
sequence.  If $\beta$ is very likely to be accessible from $\alpha$,
we shall call $\beta$ ``near'' $\alpha$.  In the case of neutrality a
shape $\alpha$ is realized by a large set of sequences, and a robust
notion of accessibility then comes to mean that $\beta$ must arise
from $\alpha$ with a high probability {\it when averaged over all
sequences folding into $\alpha$}.  Only then are the shapes in the
phenotypic neighborhood of $\alpha$ a robust property of $\alpha$
itself, independent of a particular sequence.

\begin{figure}[ht]
\vspace*{0.3cm}
\centerline{\psfig{figure=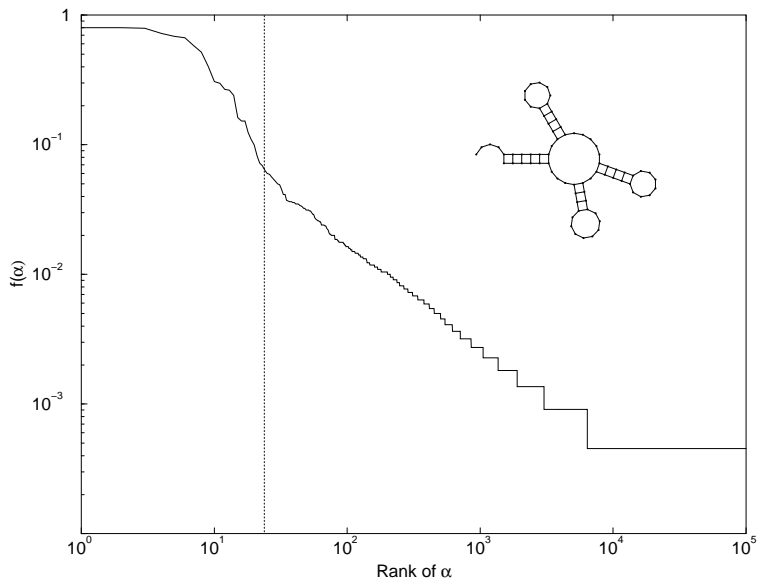,width=10cm,angle=-90}}
\caption[]{\label{zipf}{\small {\bf Rank-ordered frequency distribution
of shapes in the tRNA boundary.}  A sample of 2,199 sequences whose
minimum free energy secondary structure is a tRNA clover-leaf (inset)
was generated. All their one-error mutants (501,372 sequences) were
folded. 28\% of the mutants retained the original structure (i.e.~were
neutral).  The remaining 358,525 sequences realized 141,907 distinct
shapes. The frequency $f(\alpha)$ is the number of one-error
neighborhoods in which $\alpha$ appeared at least once, divided by the
number of sequences in the sample. The log-log plot shows the rank of
$\alpha$ versus $f(\alpha)$.  Rank $n$ means the $n$th most frequent
shape.  The dotted line indicates a change in the slope which we take
to naturally delimit the high frequency domain (to the left) whose
shapes form the characteristic set of the tRNA.}}
\vspace*{0.3cm}
\end{figure}

This notion of neighborhood is illustrated by considering a tRNA-like
shape of length 76 (inset of Figure \ref{zipf}).  A sample of the many
sequences folding into this shape was obtained by an inverse folding
procedure \cite{hofacker:94} available with the {\it Vienna RNA
package\/} \cite{package}. For every sequence in the sample we compute
all shapes realized in its sequence space neighborhood, consisting of
all $228$ one-error mutants. From this data we determine the fraction
of sequence neighborhoods in which a particular mutant shape appeared
at least once. The totality of these mutant shapes, irrespective of
how often they occurred, is termed the shape space {\it boundary\/} of
the tRNA shape.

\begin{figure}[ht]
\centerline{\psfig{figure=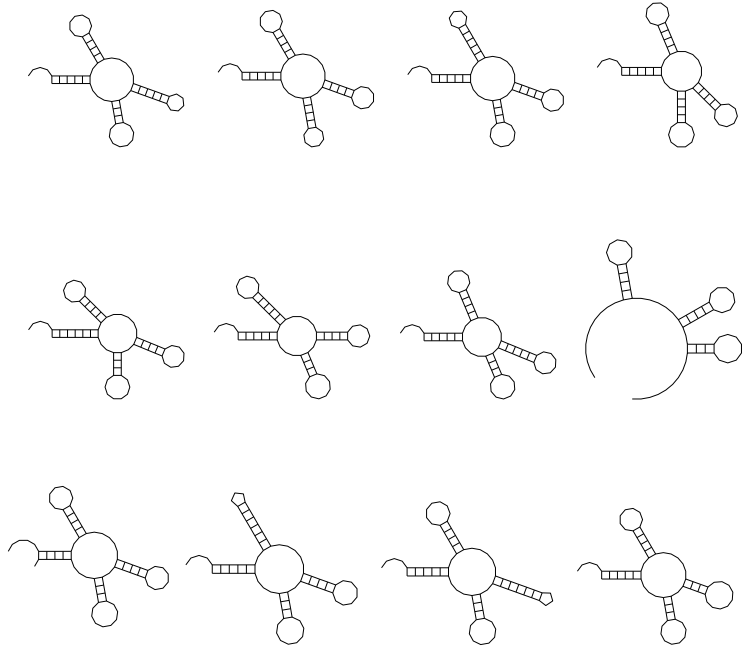,width=10cm}}
\caption[]{\label{top12}{\small {\bf Shapes near the tRNA shape.}
The figure shows the 12 highest ranked shapes (left to right, top to
bottom) in the characteristic set. See also Figure \ref{zipf}.}}
\vspace*{0.3cm}
\end{figure}

When rank-ordering the boundary shapes with decreasing frequency, we
obtain Figure \ref{zipf}. The most salient feature is a marked change
in the scaling exponent, suggesting a natural cut-off point for the
definition of neighborhood. In the present case, the high frequency
range comprises some 20 shapes, which we define to be near the tRNA
shape \cite{fontana:98b}. These shapes constitute the characteristic
set of the tRNA, that is, its most specific neighborhood.  The topmost
12 shapes are shown in Figure \ref{top12}, and exhibit two properties
we found to hold for all shapes whose neighborhoods we studied.
First, most shapes in the characteristic set of a shape $\alpha$ are
highly similar to $\alpha$, typically differing in a stack size by
single base pairs. Second, some shapes, such as tRNA$_8$ (the shape
ranked 8th in Figure \ref{top12}), differ by the loss of an entire
stack.  The latter finding illustrates that nearness of a shape to
another does not imply their similarity.  More importantly, it
illustrates that nearness is not a symmetric relation.  In fact, the
tRNA shape was not found in the characteristic set of the tRNA$_8$,
and it did not even occur in its boundary sample. Not surprisingly,
the destruction of a structural element (in a random sequence bearing
it) through a single point mutation is easier than its creation. While
the high frequency of the event is surprising, it is ultimately a
consequence of the average base pair composition of stacks and the
markedly different stacking energies of AU and GC base pairs
\cite{fontana:98b}.

The nearness relation developed here defines a topology on the set of
RNA shapes. We call a transition from shape $\alpha$ to a near shape
$\beta$ {\it continuous\/} (in the spirit of topology), if that
transition is caused by a single point mutation (since in that case
the sequence of $\beta$ is in the obvious sense near the sequence of
$\alpha$).

\begin{figure}[ht]
\centerline{\psfig{figure=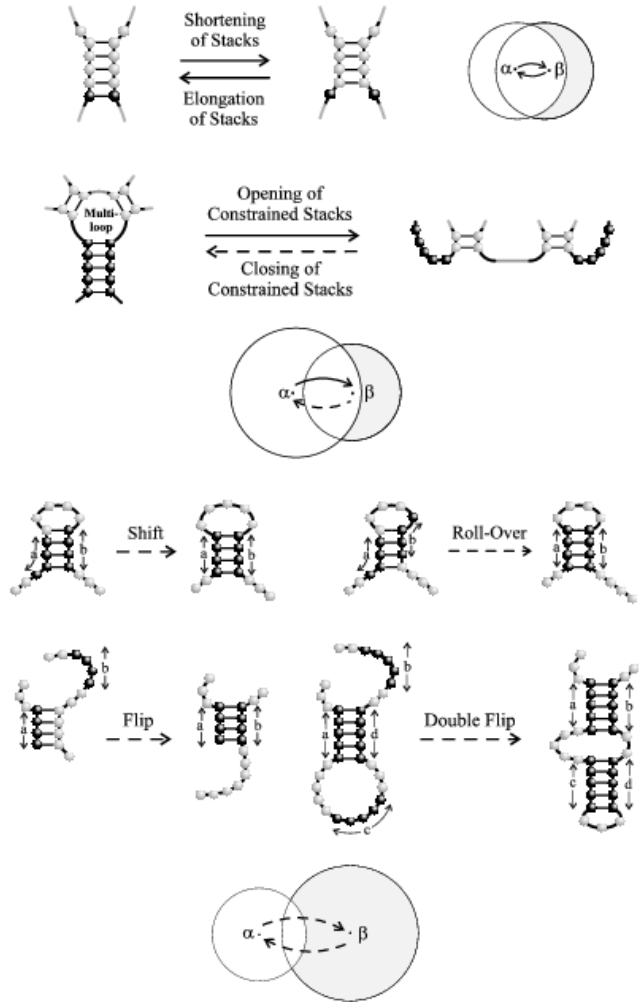,width=8.5cm}}
\caption[]{\label{trans}{\small {\bf Continuous and discontinuous
RNA shape transformations.}  The figure illustrates transformations
between RNA secondary structure parts.  Solid (dashed) arrows indicate
continuous (discontinuous) transformations in our topology. Three
groups of transformations are shown. Top: the loss and formation of a
base pair adjacent to a stack are both continuous. Middle: the opening
of a constrained stack (e.g. closing a multiloop) is continuous, while
its creation is discontinous. This reflects the fact that the
formation of a long helix between two unpaired random segments upon
mutation of a single position is a highly improbable event, whereas
the unzipping of a random helix is likely to occur as soon as a
mutation blocks one of its base pairs. Bottom: generalized shifts are
discontinuous transformations in which one or both strands of a helix
shift ending up with or without an overlap.  Accordingly, we partition
generalized shifts into the four classes shown.  The intersecting
disks are a schematic representation of continuous and discontinuous
transitions between two shapes $\alpha$ and $\beta$.  The disk with
center $\alpha$ ($\beta$) stands for the set of shapes that are near
$\alpha$ ($\beta$). If $\beta$ is a member of $\alpha$'s disk
(neighborhood), a transition from $\alpha$ to $\beta$ is continuous
(solid arrow). A discontinuous transition leaves the neighborhood of
$\alpha$ (dashed arrow).  Note that even if $\alpha$ and $\beta$ are
highly dissimilar, $\alpha$ might nontheless be transformed
continuously into $\beta$ through intermediate shapes whose
neighborhoods have sufficient overlap.}}
\end{figure}

Generalizing from this example, we can characterize continuous
transformations as those structural rearrangements which fine tune a
shape architecture in a sequential fashion by lengthening or
shortening stacks, or which destroy a stack element and the loop
implied by it (Figure \ref{trans}). This leaves the {\it
discontinuous\/} transformations characterized by the two remaining
possible structural changes: (i) the creation of a long stack in a
single step, and (ii) generalized shifts (see Figure
\ref{trans}). As an example of a shift consider one strand
of a stacked region sliding past the other by a few positions (simple
shift).  Notice here that structural similarity does not imply
nearness.  Both types of discontinuous transformations require the
synchronous participation of several bases (or base pairs) in a
fashion that {\it cannot be sequentialized on thermodynamic grounds}.
An example for a discontinuous transition of type (i) is the formation
of a multiloop (a loop issuing more than two stacking
regions). Generally, the free energy gain upon formation of a stack
must offset the free energy loss from the loop caused by it.  A stack
closing a multiloop must, therefore, come into existence with some
minimum length (typically more than 5 bp) in a single step. Likewise,
the discontinuity of generalized shifts (type ii) has thermodynamic
and structural origins. Shifting a stack by shifting its base pairs in
random order would cause energetically unfavorable internal loops as
well as severe sterical conflicts, besides violating the formal
no-knot condition (section \ref{RNAdef}).  As a consequence, the
shifting of a stack requires that all base pairs move synchronously.

\afterpage{\clearpage}

We may visualize the neighborhood structure on the set of all shapes
(the topology) as a directed graph. Each shape is represented by a
node. Directed edges fan out from a node $\alpha$ to the nodes in its
characteristic set.  We can think of a continuous transformation of
shape $\alpha$ into a shape $\beta$ that is not nearby $\alpha$ as a
path from $\alpha$ to $\beta$ following the direction of the edges.
Discontinuous transformations are transitions between disconnected
components of the graph.

\subsection{Adaptive dynamics in RNA}

We next discuss how the topology of RNA shape space shapes adaptive
histories. Our simulation of replicating and mutating RNA populations
is cast in terms of stochastic chemical kinetics
\cite{gillespie:76,gillespie:77}, and represents a continuous time model
of Spiegelman's classic serial transfer experiments
\cite{spiegelman:71,kramer:74}. It's implementation is described elsewhere
\cite{fontana:87,fontana:89}.  In the laboratory a goal might be to
find an RNA aptamer binding to some molecule
\cite{gold:90}, and the evolutionary end product is typically
unknown. In principle one can think of the end product as some shape
that is specified implicitly by the imposed selection criterion.
Since our intent is to study evolutionary trajectories rather than end
products, we short cut by simply defining a target shape in advance.
We then assume the replication rate constant of a sequence to be a
function of the distance between its shape and the target. Given a
distance measure $d$, a shape replicates faster, the more it resembles
the target.  In all simulations reported here, the replication rate
$r_i$ of a sequence I$_i$ of length $n$ with shape $\alpha$ at
distance $d(\alpha,\tau)$ from a target shape $\tau$ is given by
$r_i=(0.01 + d(\alpha,\tau)/n)^{-1}$.  Using an exponential or a
linear function did not make any difference with regard to the issues
we are interested in. The error rate was $p=0.001$ per nucleotide.  At
this rate, the difference between parent and a mutant offspring is
mostly one point mutation, and the topology described above applies.
In the examples reported here the target shape is a tRNA clover leaf,
and the distance between shapes is measured as the Hamming distance
between their line oriented representations cast in terms of dots and
parentheses (see section \ref{RNAdef}).

\begin{figure}[ht]
\centerline{\psfig{figure=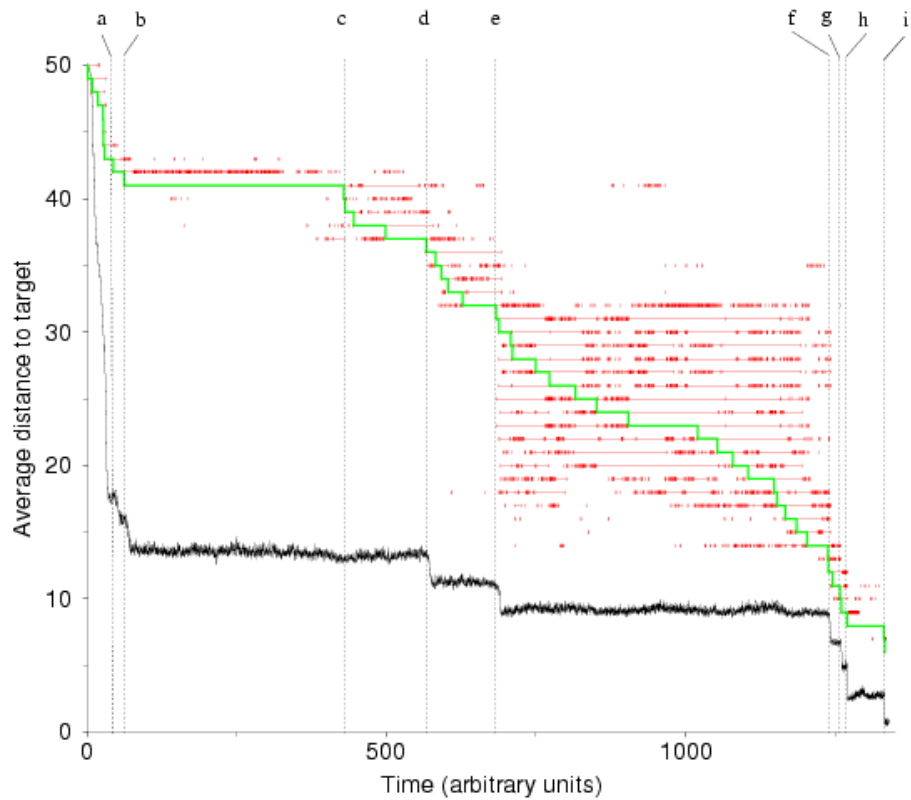,width=12cm,angle=-90}}
\caption[]{\label{evolve1}{\small {\bf Simulation of an evolving
RNA population.}  An RNA population evolves towards a tRNA target
shape (inset of Figure \ref{zipf}) in a flow reactor logistically
constrained to a capacity of 1,000 sequences on average. The
replication accuracy per position is $0.999$.  The initial population
consisted of 1,000 identical sequences folding into a random shape.
The target was reached after approximately $11\times 10^6$
replications. The black trace shows the average structure distance of
the shapes in the population to the target.  The relay series (see
text) linking the initial shape to the target comprises 43 shapes.  To
each of these corresponds one horizontal level placed above the black
curve. The topmost (bottom) level belongs to the initial (target)
shape. For these levels only the time axis has a meaning.  At each
level the series of presence intervals for the corresponding shape is
shown. The step curve indicates the transitions between relay shapes,
and hence the time spent by each relay shape on the adaptive
trajectory. Each transition was caused by a single point mutation in
the underlying sequences. The vertical dotted lines and the labels
mark transitions referred to in the text.}}
\vspace*{0.3cm}
\end{figure}

\begin{figure}[ht]
\centerline{\psfig{figure=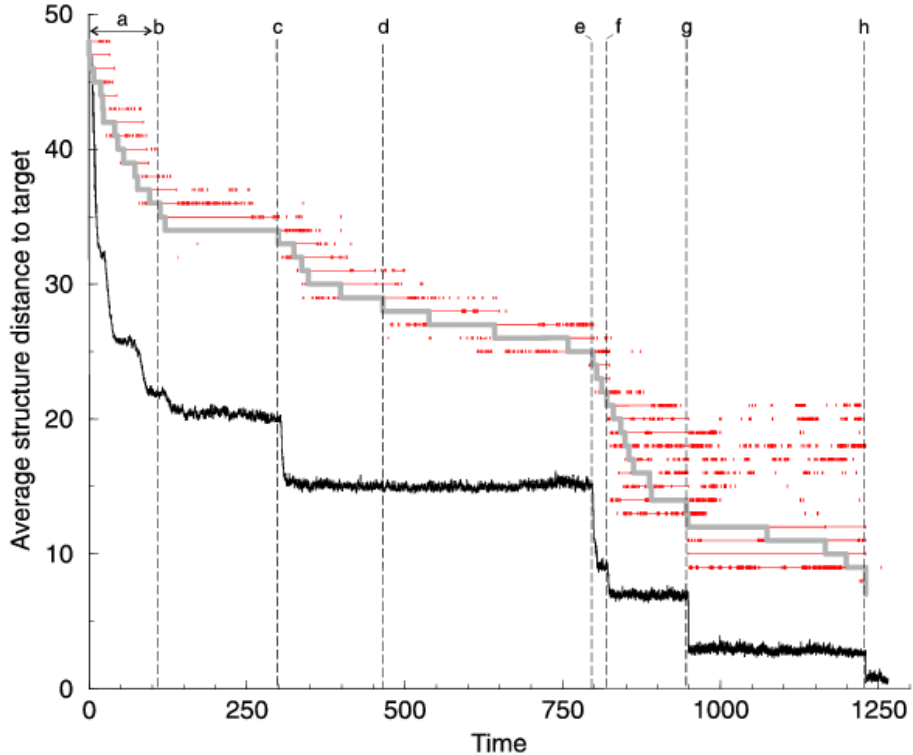,width=12cm}}
\caption[]{\label{evolve2}{\small {\bf Simulation of an evolving
RNA population.}  Same situation as in Figure \ref{evolve1}, except
for a different initial random shape. The labels refer to shape
transformations shown in Figure \ref{relay2}}}
\vspace*{0.3cm}
\end{figure}

Figure \ref{evolve1} and Figure \ref{evolve2} both show the approach
toward the target shape as indicated by the average Hamming distance
in the population (inversely related to fitness, black curve).  Aside
from a short initial phase, the entire history is dominated by steps,
that is, flat periods of no apparent adaptive progress, interrupted by
sudden improvements toward the target. The initial relaxation period
is understood by considering that many modifications of a random
initial shape increase its similarity to any randomly chosen
target. Beyond this phase, however, adaptation becomes much harder,
and the character of adaptive dynamics changes.

\begin{figure}[ht]
\centerline{\psfig{figure=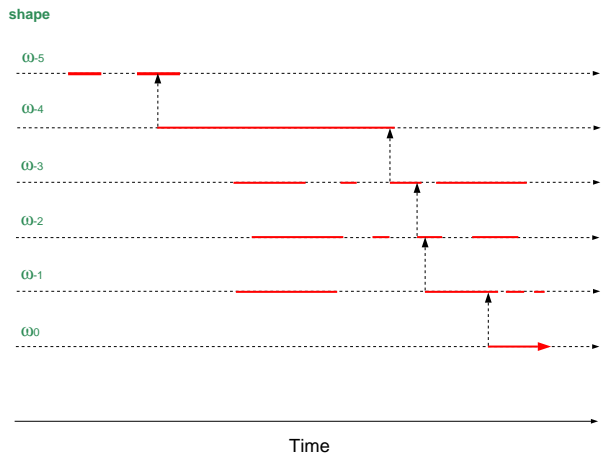,width=8cm}}
\caption[]{\label{relay}{\small {\bf Relay series concept.}
The figure shows schematically the presence intervals of 6 shapes, and
the reconstruction of the uninterrupted chain of innovations called
the relay-series. See text for details.}}
\vspace*{0.3cm}
\end{figure}

By adaptive trajectory we mean something like a path taken by the
population as a whole (caveats below), rather than a single lineage.
We refer to its projection on shape space as its phenotypic trace. In
a simulation of this kind one has to cope with a huge amount of data,
and one possibility of obtaining an approximation to the phenotypic
trace of an adaptive trajectory is to record only data about mutation
events that generate an ``innovation'', that is, a shape which is new
in the population at the time $t$ of its appearance. This does not
neccesarily imply that the shape hasn't been in the population in the
past; it could have become extinct at some earlier time, being
``rediscovered'' at time $t$.  For each shape innovation $\alpha$ we
record entry times, $l_i^{\alpha}$, and exit (extinction) times
$h_i^{\alpha}$. This yields for each shape ever seen during the
adaptive process a set of ``presence intervals''
${L}_{\alpha}=\{[l_i^{\alpha}, h_i^{\alpha}],\;l_i^{\alpha} <
h_i^{\alpha} < l_{i+1}^{\alpha}\}$, marking the presence of shape
$\alpha$ in the system's history.  After the target has been found (or
the simulation has been stopped), we trace back through the history
data in the following way.  Each presence interval $[l_i^{\alpha},
h_i^{\alpha}]$ of $\alpha$ has a unique ancestor with shape $\beta$
which spawned that interval at time $l_i^{\alpha}$, meaning that a
sequence folding into $\beta$ produced at time $l_i^{\alpha}$ a mutant
which folded into $\alpha$, and $\alpha$ was not in the population at
that time. Let $\omega$ be the target shape, and $l_i^{\omega}$ the
time of its first appearance (the beginning of its presence interval).
Let the shape from which $\omega$ was derived at time $l_i^{\omega}$
be $\omega_{-1}$.  In the set ${L}_{\omega_{-1}}$ there is a unique
presence interval $[l_j^{\omega_{-1}}, h_j^{\omega_{-1}}]$ containing
the time instant $l^{\omega}_i$, and we proceed searching for the
unique ancestor of $[l_j^{\omega_{-1}}, h_j^{\omega_{-1}}]$. Upon
repeating this procedure we eventually end up at one of the initial
shapes (see Figure \ref{relay}). At this point we have reconstructed a
succession of shapes
$\alpha\equiv\omega_{-n}\;\omega_{-n+1}\;\cdots\;\omega_{-i}\;
\cdots\;\omega_{-1}\;\omega_{0}\equiv\beta$ connecting an initially
present shape $\alpha$ with the target (or final) shape $\beta$. This
chain is uninterrupted in time, in the sense that for every $n\ge i\ge
1$, $\omega_{-i}$ is ancestor of $\omega_{-i+1}$ and there exists a
pair $[l_r^{\omega_{-i}}, h_r^{\omega_{-i}}]\; [l_s^{\omega_{-i+1}},
h_s^{\omega_{-i+1}}]$ with $l^{\omega_{-i}}_r < l^{\omega_{-i+1}}_s <
h_r^{\omega_{-i}}$.  The chain depends on the presence interval of the
final shape $\beta$ from where the trace starts, but it is unique for
that interval.  Typically we are interested in the chain that
originates in the first instance of the target, and call it the
``relay series''.

\vspace*{0.3cm}

\begin{figure}[ht]
\centerline{\psfig{figure=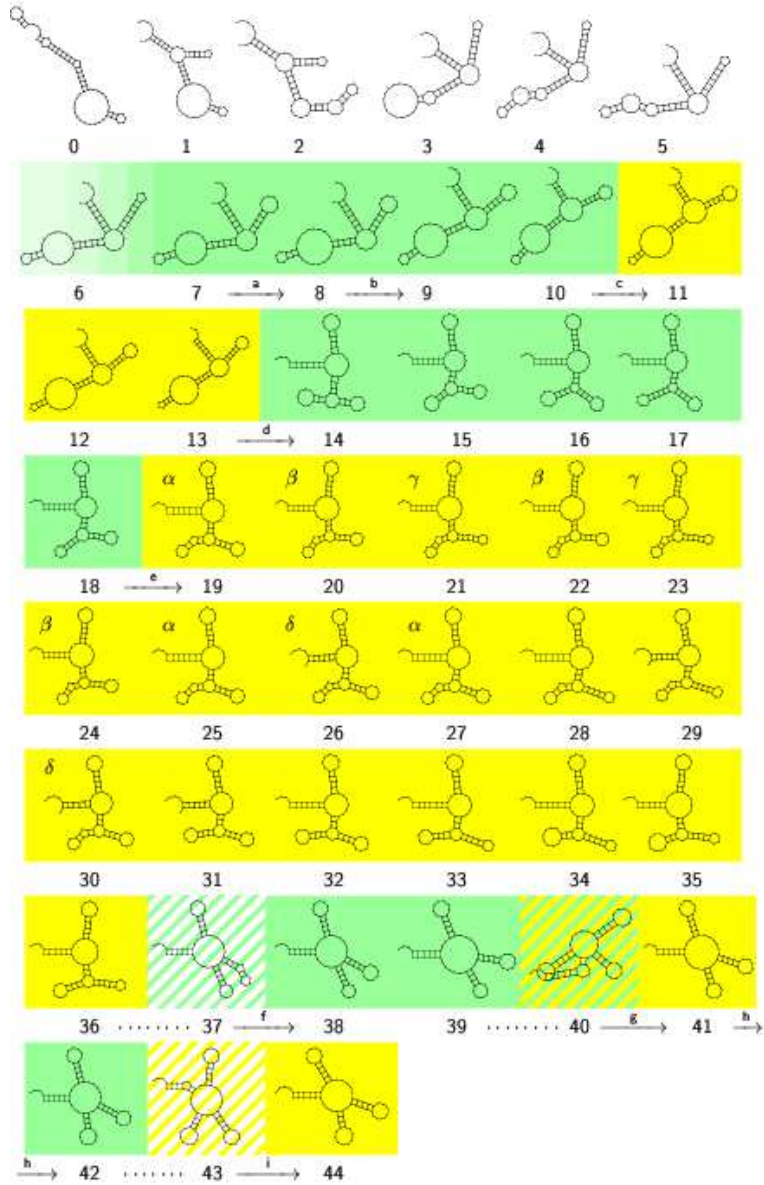,width=10cm}}
\caption[]{\label{relay1}{\small {\bf Relay series of Figure \ref{evolve1}.}
The full series of 45 relay shapes is shown. Different gray levels indicate
different classes of neighboring shapes which are accessible from the
precursor class by a discontinuous transition only. The stretch 19 to 
37 (corresponding to the long plateau e-f in figure \ref{evolve1}) 
shows a sequence of continuous transformations between closely related 
shapes some of them occurring more than once in the relay series: 
$\alpha\to\beta\to\gamma\to\beta\to\gamma\to\beta\to\alpha\to\delta\to\alpha$.
}}
\vspace*{0.3cm}
\end{figure}

\begin{figure}[ht]
\centerline{\psfig{figure=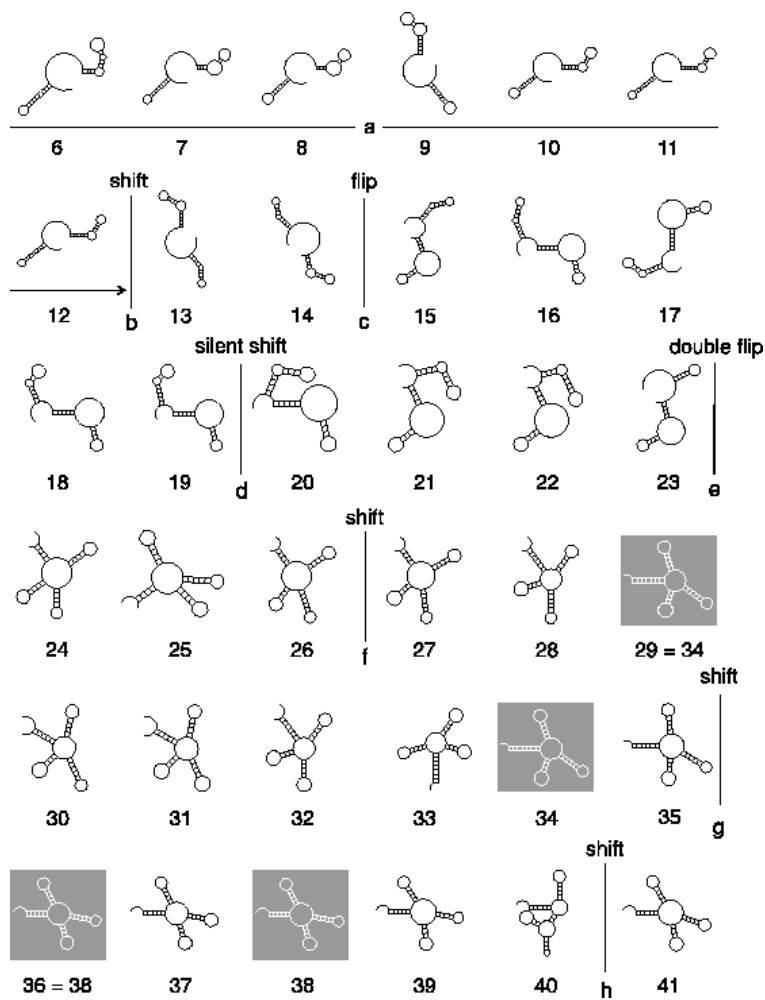,width=10cm}}
\caption[]{\label{relay2}{\small {\bf Relay series of Figure \ref{evolve2}.}
This series involves 42 relay shapes. Discontinuous transitions 
classified as in Figure \ref{trans} are marked by letters referring
to Figure \ref{evolve2}. Again, shapes occur repeatedly in the series
as indicated, for example, by the white shapes on gray background.}}
\vspace*{0.3cm}
\end{figure}

Figures \ref{relay1} and \ref{relay2} show the complete relay series
for the computer simulations reported in Figures \ref{evolve1} and
\ref{evolve2}, respectively. While the discontinuities in the fitness
trace of Figures \ref{evolve1} and \ref{evolve2} are apparent, it is
their comparison with the shape discontinuities along the relay series
which yields insight. The result is that fitness triggered selection
events do not always line up with shape transformations that are
discontinuous in the sense of the previously defined topology. In
Figure \ref{evolve1}, events (a) and (b) are rapid successions of
continuous transitions shortening and elongating stacks by single base
pairs. This shows that sudden changes in fitness do not imply
discontinuous phenotypic transformations. The reverse isn't true
either, as shown by the discontinuous shift event (c) which is silent
in terms of fitness. All remaining fitness changes do, however,
coincide with discontinuous shape transformations. These are the
simple shift events (e), (g), (h), (i), the double flip (d), and the
flip (f). Similar observations hold for Figure \ref{evolve2}. Here we
have an initial phase of rapid improvments (a), four simple shift 
events (b), (f), (g), (h), a flip (c), a double flip (e) as well as
a ``silent shift'' (d) being a neutral discontinuous transitions, 
that is, a shift which does not change fitness.

To trigger a generalized shift or a multiloop closure by a single
point mutation puts constraints on the required sequence context,
making such sequences rare. When a shape $\alpha$ is under strong
selection, neutral drift is the only means for eventually producing
such a sequence \cite{huynen:96a,huynen:96c}.  This is why
discontinuous transitions are preceded by extended periods of neutral
drift in Figures \ref{evolve1} and \ref{evolve2}.

It is important to realize that the present reconstruction of an
adaptive trajectory is not identical to the succession of shapes
associated with the actual lineage of sequences that led to the
target.  The relay series reports for all $n$ from start to target
that (some sequence folding into) shape $\omega_n$ was sufficient for
giving rise to a new shape $\omega_{n+1}$. In the actual successful
lineage $\omega_{n+1}$ may not have arisen from the first sequence
with shape $\omega_n$; for that matter, it may not even have arisen
from any sequence with shape $\omega_n$.  (Sequences with other shapes
may also have produced $\omega_{n+1}$, while $\omega_{n+1}$ was
already present in the population.)  In sum, the relay series is a
{\it fictitious\/} path. The point is that it is a {\it possible\/}
path and a very useful representative of the {\it ensemble\/} of paths
shown to be possible by a population in a particular adaptive history.
Let us explain.  At any given time $t$ a population contains a number
of individuals (instances) for each of the sequence types
present. Each individual has its unique lineage of ancestors all the
way back to the initial individuals. At the level of individuals there
would obviously be only one successful lineage or causal path to the
first appearance of the target.  However, it makes sense not to
distinguish among identical individuals, and to think in terms of
sequence types (or molecular ``species'') instead. The consequence is
that a given sequence type can be produced by more than one other
sequence type, thus giving rise to a network.  At this level of
resolution there is a combinatorics of paths relating the first
sequence type possessing target shape to the sequence types present
intially. Although we have lost strict causality, each of these paths
could have been a causal one if the dice had rolled differently. We
hence refer to this set as the ``ensemble of paths shown to be
possible by a population in a particular history''.  To each path in
this ensemble corresponds a phenotypic trace, and the relay series is
one of those.  Now comes the salient point. All these different paths
must coincide at least at the discontinuous transitions revealed by
the relay series. These transitions are seen, after the fact, as those
key innovations that enabled the population to reach the target in a
particular experiment. If such a transition is associated with a
fitness advantage, then we have a typical founder effect, where one
(or a few related) sequences conveying this advantage are amplified by
selection, giving rise to all future lineages, while the other
lineages up to that epoch are terminated. If a discontinuous
transition is not associated with a fitness advantage we have what we
might call a silent founder effect. All lineages other than those
emanating from the silent founder are doomed at some later point,
since only a lineage from the silent founder enables the next major
transition. In other words, discontinuities in the relay series
indicate way points in shape space where ensembles of different paths
coincide as they zero in (even in the absence of fitness advantage) on
the target.  The few paths reaching such a way point diversify
afterwards on shape neutral as well as fitness neutral networks, until
they are reduced again at the next way point. To be sure, any fitness
improving transition has this reducing effect, whether it is
associated with a discontinuous transformation or not. The important
point here is that fitness improving transitions can be anything
depending on the selection criterion, and hence cannot be
characterized in general. In contrast, the characterizable class of
discontinuous shape transformations is an intrinsic property of RNA
folding, and constitutes a set of potential way points for {\it any\/}
selection criterion.

Preliminary results from simulations with different population sizes
(same initial conditions, same target) show roughly a constant number
of discontinuous transitions for different adaptive trajectories,
while consistently reducing the number of continuous transitions.

To summarize, the emerging picture of an adaptive trajectory is not
one of a single path, but rather of an ensemble of diverse paths
coming together at way points characterizable as discontinuous shape
transformations.  (This ``ensemble view'' contains imagery borrowed
from \cite{dill:97,nimw97a}.)  Given a particular initial condition
and a target, different simulations yield different adaptive
trajectories. However, all involve the same class of way points.
Future work will have to track several individual lineages throughout
the adaptive process and ``align'' them to corroborate this picture. A
hint in this direction is provided by the pattern of presence
intervals of relay shapes during the history of a simulation (see
Figures \ref{evolve1} and \ref{evolve2}). These patterns nicely
visualize the nearness relation between shapes and the way point
concept. When a shape $\alpha$ is succeeded by a shape $\beta$ that is
near $\alpha$, $\beta$ is present intermittently in the population
well prior to becoming part of the relay path. That is, once $\alpha$
is present, $\beta$ is unavoidable, and a transition to $\beta$ is
continuous. Conversely, at a discontinuous transition, when $\alpha$
is succeeded by a shape $\beta$ that is not near $\alpha$, $\beta$ has
almost always its first ever appearance just prior to that
transition. Seen together, the presence intervals of successive shapes
on the relay path form blocks of continuous (within-neighborhood)
transitions, separated by discontinuous transitions (neighborhood
escapes). During within-neighborhood transitions many relay shapes
coexist and the ensemble of possible paths is distributed over these
shapes.

\section{Conclusions}

We have reviewed our theoretical understanding of evolutionary
adaptation in model populations of RNA sequences subject to selection
at the level of their secondary structures. The focus was on the
evolutionary consequences of neutrality. The issue of neutrality in
estimating the rate of evolution was brought to the fore by the
Japanese population geneticist Motoo Kimura in the late sixties
\cite{kimura:68,kimura:83}, and was triggered by the observation of
high aminoacid substitution rates in proteins. Neutrality was recently
emphasized again by Sergey Gavrilets in the context of model
landscapes \cite{gavrilets:97}, so-called ``holey adaptive
landscapes'' consisting of neutral networks with a ``swiss
cheese''-like structure.  The extreme case of stochastic dynamics on a
flat landscape was approached by Peliti and Derrida
\cite{derrida:91}, while recent progress focussed on stochastic
adaptive dynamics in piece-wise neutral prototype landscapes, such as
landscapes of a wedding-cake structure \cite{nimw98a,nimw98b,nimw96a},
``neutralized'' NK-type (Kauffman) landscapes
\cite{barnett97,newman:98}, or landscapes based on random graphs
\cite{pks:248}.  The term neutral network was coined in
\cite{schuster:94a} where the phenomenon was found in the context of
the RNA folding map.

The interest in RNA secondary structure folding derives from providing
a compromise between landscapes designed for analytic tractability and
landscapes grounded in molecular reality that are also suited as
laboratory models \cite{biebricher:97,joyce:93}. Our findings on RNA 
folding provide a microfoundation for Kimura's phenomenological approach, 
and led to insights which are hard to obtain without a mechanistic model.  
We briefly summarize these insights.

First, the important fact about neutrality in RNA is that sets of
shape neutral sequences are connected by single point mutations, and,
hence, are not just sets but networks. This alone means that adaptive
populations can tolerate higher mutation rates for transmitting a
master phenotype over generations, than if a particular genotype had
to be transmitted.

Second, while neutrality means robustness against mutations, it also
means an increased {\it ability to adapt\/}. This is less paradoxical
than it may look at first sight. The effect of a mutation depends on
the context in which it is expressed. By permitting the sequence
context to vary while preserving a shape, neutrality is a prerequisite
for making certain point mutations consequential, and hence enables
phenotypic innovation. Phrased in terms of landscape vocabulary,
neutral networks delocalize ``local traps'', and change the way a
``trap'' works. In a barrier-like trap nothing happens for a long
time, while the system is waiting for an improbable
macromutation event (such as several point mutations at once) to
occur. In contrast, a neutral network is a diffusion-like trap in
which the population spends time drifting in sequence space (as well
as in fitness neutral parts of shape space) until it hits some rare
region of the network where an adaptive transition to some other
network becomes possible. Intuitively, converting a barrier-like trap
into a (comparable) diffusion-like trap increases the likelihood of a
transition, but analytic calculations are required to spell out the
exact conditions for this to be the case \cite{nimw:pers}. (The
terminology of a barrier-like and diffusion-like trap is derived from
the similar notions of an energy barrier and an entropic barrier in
entropic spin-glasses \cite{newman:pers}.) The importance of extended
neutral networks is nicely illustrated by our repeated failure to
evolve in computer simulations a tRNA shape with {\bf GC}-only
sequences. {\bf GC}-only sequences with tRNA shape do exist, since we
could obtain thousands of them by inverse folding. Yet, each simulated
adaptive process starting from a random initial condition would get
stuck far away from any tRNA shape. Structure landscapes based on
binary {\bf GC}-only sequences are very rugged \cite{fontana:93a}, and
do not have sufficient neutrality to remove the barrier-like nature of
local traps
\cite{schuster:94a}. We had the same failure in the case of {\bf
AU}-only sequences. However, we were unable to find any {\bf AU}-only
sequence with tRNA shape by inverse folding, and the failure to evolve
one might well be due to its non-existence.

Third, neutral networks suggest a topology for RNA shapes reflecting
their mutual ``accessibility''. Accessibilty (by  point mutation) of one
shape from another depends on the fraction of boundary shared by their
corresponding neutral networks. This is not a symmetric relation.  A
random step out of Monaco has a very high probability of ending in 
France, but the reverse is not true.  In this topology a transition
from shape $\alpha$ to $\beta$ might be rare along the direct route,
but feasible in an incremental fashion by chaining together frequent
transitions. However, we found that there exist rare transitions whose
likelihood cannot be increased by any indirect route, and, hence, they
are irreducibly rare. These transitions are discontinuous in a
topological sense, and we did characterize that set as those structure
transformations which require the coordinated change of several parts
of the molecule at once. This characterization may apply to systems
other than RNA as well.  The fact that for RNA that set also has a
thermodynamic raison d'\^etre, nicely illustrates how the physics of
biopolymer structure constrains adaptive trajectories quite
independently of external fitness criteria.

Fourth, neutrality increases the uncertainty in the genotypic trace of
adaptive trajectories. However, their phenotypic trace must go through
discontinuous transitions, no matter what the target shape is made to
be. This again reduces to some extent the uncertainty of adaptive
trajectories at the phenotypic level.

The same punctuated dynamics that we have observed in RNA, was also 
found with evolving bacteria under precisely controlled constant 
conditions in chemostats \cite{elena:96} stressing once more its 
intrinsic nature. Punctuation has been reported as well in a quite 
different context where cellular automata are
evolved to perform specific computational tasks (yet another kind of
``genotype-phenotype'' map) \cite{crutchfield:94a,crut98c}. This
raises the possibility that neutral networks giving rise to a
punctuation dynamics are a quite general phenomenon, not limited to
materials or structures related to strictly biological systems.

It is thought provoking to consider neutrality as a function of the
resolution at which we conceptualize ``shape'' (or ``structure'' in
general). As we decrease the resolution, formely distinct neutral
networks will merge, and diversity is lost until it makes no sense to
speak of an adaptive process, since the features captured by that
level of resolution have all the same adaptive value, that is, none.
On the other hand, as we increase the level of resolution, a formerly
single neutral network splits into smaller networks. In view of the
importance we ascribe to neutrality for adaptability, it becomes
meaningful to ask at which level of resolution networks become so
confined as to destroy the adaptability of a system. It is doubtful,
for example, whether RNA could be evolvable at all, if the full set of
atomic coordinates of an RNA shape were to matter for its function.
In that case no two sequences were structurally the same. This train
of thought suggests that the first appearance of percolating
neutrality on the structure resolution scale defines
the evolutionary relevant notion of ``structure''.

\newpage
\vspace{0.3 cm}
{\bf Acknowledgements}

Financial support of the work presented here was provided by the
Austrian {\em Fonds zur F\"orderung der wissenschaftlichen Forschung}
(Projects P-11\,065 and P-13\,093), by the Commission of the European 
Union (Project PL-970\,189), and by the Santa Fe Institute.

\vspace{0.3 cm}
\bibliography{predict}
\bibliographystyle{uns-abbr}
\end{document}